\documentclass[aps,twocolumn,superscriptaddress,floatfix,preprintnumbers,nofootinbib]{revtex4-2}
\usepackage[utf8]{inputenc}
\usepackage{amsmath}
\usepackage{amssymb}

\usepackage{enumerate}
\usepackage{amsmath}
\usepackage{tikz}
\usepackage[compat=1.1.0]{tikz-feynman}
\usepackage{feynmf}
\usepackage{slashed}
\usepackage{braket}
\usepackage[toc,page]{appendix}
\usepackage{url}
\usepackage{natbib}
\usepackage{graphicx}
\usepackage[pdftex,bookmarks,linktocpage,pdfpagelabels,plainpages=false,hyperfigures,linkcolor=blue,citecolor=blue]{hyperref} 
\hypersetup{colorlinks=true}

\usepackage{mathrsfs,amssymb}  
\usepackage{cancel}
\usepackage[normalem]{ulem}

\begin{document}
\begin{flushright}
MI-TH-2135
\end{flushright}

\author{P.~S.~Bhupal Dev}
\email{bdev@wustl.edu}
\affiliation{Department of Physics and McDonnell Center for the Space Sciences, Washington University, St. Louis, MO 63130, USA}

\author{Doojin~Kim}
\email{doojin.kim@tamu.edu}
\affiliation{Mitchell Institute for Fundamental Physics and Astronomy,
Department of Physics and Astronomy, Texas A\&M University, College Station, TX 77843, USA}

\author{Kuver Sinha}
\email{kuver.sinha@ou.edu}
\affiliation{Department of Physics and Astronomy, University of Oklahoma, Norman, OK 73019, USA}

\author{Yongchao Zhang}
\email{zhangyongchao@seu.edu.cn}
\affiliation{School of Physics, Southeast University, Nanjing 211189, China}
\affiliation{Department of Physics and McDonnell Center for the Space Sciences, Washington University, St. Louis, MO 63130, USA}

\title{PASSAT at Future Neutrino Experiments: \\
Hybrid Beam-Dump-Helioscope Facilities to Probe Light Axion-Like Particles
}

\begin{abstract}
There are broadly three channels to probe axion-like particles (ALPs) produced in the laboratory: through their subsequent decay to Standard Model (SM) particles, their scattering with SM particles, or their subsequent conversion to photons.
Decay and scattering are the most commonly explored channels in beam-dump type experiments, while conversion has typically been utilized by light-shining-through-wall (LSW) experiments. 
A new class of experiments, dubbed PASSAT (Particle Accelerator helioScopes for Slim Axion-like-particle deTection), has been proposed to make use of the ALP-to-photon conversion in a novel way: ALPs, after being produced in a beam-dump setup, turn into photons in a magnetic field placed near the source. 
It has been shown that such hybrid beam-dump-helioscope experiments can probe regions of parameter space that have not been investigated by other laboratory-based experiments, hence providing complementary information; in particular, they probe a fundamentally different region than decay or LSW experiments. 
We propose the implementation of PASSAT in future neutrino experiments, taking a DUNE-like experiment as an example.
We demonstrate that the magnetic field in the planned DUNE multi-purpose detector is already capable of probing the ALP-photon coupling down to $g_{a\gamma\gamma} \sim {\rm few}\times 10^{-5}$ GeV$^{-1}$ for ALP masses $m_a \lesssim 10$ eV. The implementation of a CAST or BabyIAXO-like magnet would improve the sensitivity down to $g_{a\gamma\gamma} \sim 10^{-6}$ GeV$^{-1}$.

\end{abstract}

\maketitle

\section{Introduction} \label{sec:intro}
Motivated by the strong CP problem, QCD axions have been proposed~\cite{Peccei:1977hh,Weinberg:1977ma,Wilczek:1977pj} and experimentally searched for  over the past several decades~\cite{Jaeckel:2010ni, Irastorza:2018dyq}. 
Meanwhile, the theoretical and experimental efforts have been extended to incorporate more general pseudoscalars, collectively called axion-like particles (ALPs). 
As the Standard Model (SM) does not contain the QCD axion nor any ALP, their discovery will be an unambiguous indication of new physics beyond the SM which may also address other unresolved problems, such as dark matter~\cite{Preskill:1982cy,Abbott:1982af,Dine:1982ah,Duffy:2009ig,Marsh:2015xka,Battaglieri:2017aum}. 
As such, a tremendous amount of experimental effort has been devoted in the search for ALPs, utilizing their couplings to photons, electrons, nucleons, or some combinations thereof. 
These experiments include helioscopes~\cite{Zioutas:1998cc,Irastorza:2013dav,Armengaud:2019uso},  
haloscopes~\cite{Asztalos:2001tf,Kahn:2016aff,JacksonKimball:2017elr, Salemi:2019xgl}, light-shining-through-wall (LSW) experiments~\cite{Ehret:2009sq,  Bahre:2013ywa, Ballou:2015cka, Spector:2019ooq},
interferometries \cite{Melissinos:2008vn,DeRocco:2018jwe,Obata:2018vvr, Liu:2018icu}, accelerator-based searches~\cite{Alekhin:2015byh,Dobrich:2015jyk,Berlin:2018pwi,Feng:2018noy,Berlin:2018bsc,Akesson:2018vlm,Volpe:2019nzt,Dusaev:2020gxi,Banerjee:2020fue,Kelly:2020dda,Brdar:2020dpr}, hybrids of beam-dump and helioscope approaches~\cite{Bonivento:2019sri}, reactor-based searches~\cite{Chang:2006ug,Dent:2019ueq,AristizabalSierra:2020rom}, dark matter experiments~\cite{Fu:2017lfc,Aralis:2019nfa, Aprile:2020tmw,Dent:2020jhf}, and emission from neutron stars/magnetars/mergers~\cite{Fortin:2018ehg, Fortin:2018aom, Lloyd:2020vzs, Harris:2020qim, Fortin:2021sst}. 

Among all these experimental endeavors, searches based on laboratory-produced ALPs (e.g., particle accelerator/reactor-based experiments and LSW-type experiments) are particularly interesting, as they can probe ALP parameter space in the most model-independent, hence conservative fashion.  
By contrast, the constraints from searches for ALPs coming from astrophysical sources (e.g., the Sun, horizontal-branch stars, red giants, and supernovae) and cosmology arguments rely on various model assumptions, which can be relaxed or even made irrelevant, depending on the underlying model details. Evasion mechanisms of stellar bounds have been widely explored in the literature, many of them in the wake of the PVLAS anomaly~\cite{Zavattini:2005tm} that could have been explained by ALPs, but only in a region of parameter space already excluded by contemporaneous CAST data~\cite{Zioutas:2004hi}. The proposed mechanisms typically introduce features in the hidden sector that switch off Primakoff production in stellar environments but restore them elsewhere: chameleon-like screening effects~\cite{Brax:2007ak}, phase transitions~\cite{Mohapatra:2006pv,Masso:2006gc}, and choices of couplings to facilitate trapping~\cite{Jain:2005nh} (a nice review of these and other studies can be found in Ref.~\cite{Jaeckel:2006xm}). In the wake of the recent EDGES~\cite{Bowman:2018yin} and Xenon1T~\cite{Aprile:2020tmw} anomalies, such model-building efforts have acquired a new urgency;  we refer to Refs.~\cite{Bloch:2020uzh, Budnik:2020nwz, DeRocco:2020xdt} for recent ideas in these directions. 

Proposals to probe ALPs in the laboratory, sometimes squarely within regions purportedly ruled out by astrophysics, have also gained steam~\cite{Bonivento:2019sri,Dent:2019ueq}. It is therefore increasingly important to (re-)explore the ALP parameter space by producing and detecting ALPs in the laboratory, even if it is already constrained by astrophysical and/or cosmological searches.

There are broadly three ways to probe relativistic ALPs in the laboratory: through their decay to SM particles, through their scattering on SM particles, and through their conversion to photons. Decay is the most widely studied~\cite{Kelly:2020dda,Brdar:2020dpr,Feng:2018noy,Berlin:2018bsc,Akesson:2018vlm,Volpe:2019nzt,Dusaev:2020gxi,Banerjee:2020fue,Berlin:2018pwi,Alekhin:2015byh,Dobrich:2015jyk}, scattering has led to recent proposals at neutrino experiments~\cite{Brdar:2020dpr,Dent:2019ueq,AristizabalSierra:2020rom}, and conversion has typically been utilized by LSW experiments~\cite{Spector:2019ooq}. 
A new class of experiments, called Particle Accelerator helioScopes for Slim Axion-like-particle deTection, or PASSAT as shorthand~\cite{Bonivento:2019sri}, has been proposed to utilize the conversion mechanism in a novel way: ALPs, after being produced by a particle beam on a target, convert to photons in a magnetic field placed near the source. 
Such hybrid beam-dump-helioscope experiments have been shown to probe regions of parameter space that have not been probed by other laboratory-based experiments; in particular, they can probe a fundamentally different region than decay or LSW experiments (see Fig.~\ref{fig:expsense}).

More specifically, if the ALP created at the target enters a region with a transverse magnetic field, then a beam-dump experiment can become sensitive to (very) light ALPs.
This is the reason why the decay of ALPs is no longer necessary; rather, the ALP \textit{converts} to a photon which then can be detected. If the length that the ALP passes through is shorter than the associated oscillation length, the conversion becomes coherent and a net conversion probability can be determined by the ALP-photon coupling. 
In practice, PASSAT enables us to explore the region of parameter space even beyond the LSW limits (governed by the laser intensity and energy) up to the region where the decay probability of ALPs becomes significant. We will explicitly demonstrate this complementarity feature in Sect.~\ref{sec:result}.

In this paper, we propose the implementation of the  PASSAT idea in neutrino-beam experiments, taking a Deep Underground Neutrino Experiment (DUNE)-like experiment as a concrete example and assuming that the ALPs couple to the SM photon. 
In neutrino-beam experiments, intense proton beams striking a target can generate not only a large number of neutrinos but a high-intensity photon flux.  An enormous number of photons emerge from bremsstrahlung in addition to meson decays, and can interact with nuclei in the target material and turn into ALPs via the Primakoff effect in the forward region.
Therefore, PASSAT implemented in the neutrino beam experiments can achieve competitive sensitivity to laboratory-produced ALP signals. 
We will make use of a full photon flux available in the target simulated by \texttt{GEANT4}~\cite{Agostinelli:2002hh}, a dedicated detector-level Monte Carlo code package, in our study. We find that the magnetic field in the proposed DUNE multi-purpose detector (MPD)  
is already capable of exploring the ALP-photon coupling down to $g_{a\gamma\gamma} \sim {\rm few}\times 10^{-5}$ GeV$^{-1}$ for ALP masses $m_a \lesssim 10$ eV. The implementation of a CAST or BabyIAXO-like magnet would improve the sensitivity down to $g_{a\gamma\gamma} \sim 10^{-6}$ GeV$^{-1}$, as shown in Fig.~\ref{fig:expsense}.

To convey our idea efficiently, the rest of this paper is organized as follows. In Sect.~\ref{sec:idea}, we briefly review the idea of PASSAT and propose two schemes of implementing PASSAT in future neutrino experiments, taking DUNE as a concrete example. We then discuss how to calculate the expected ALP signal rate at a given PASSAT system in Sect.~\ref{sec:rate}.
Our background considerations appear in Sect.~\ref{sec:background}. We then describe the simulation method that we use and present our main results in Sect.~\ref{sec:result}. Sect.~\ref{sec:conclusion} is reserved for conclusions and perspectives. Some calculation details relevant for the ALP production are relegated to Appendix~\ref{app:coordinates}.

\section{Main Idea \label{sec:idea}}

We first review the main idea of PASSAT proposed in Ref.~\cite{Bonivento:2019sri}, and then discuss ways of implementing it in DUNE-like neutrino-beam experiments. 

\subsection{PASSAT and Photon Sources}
\begin{figure}[t]
\centering
\includegraphics[width=8.4cm]{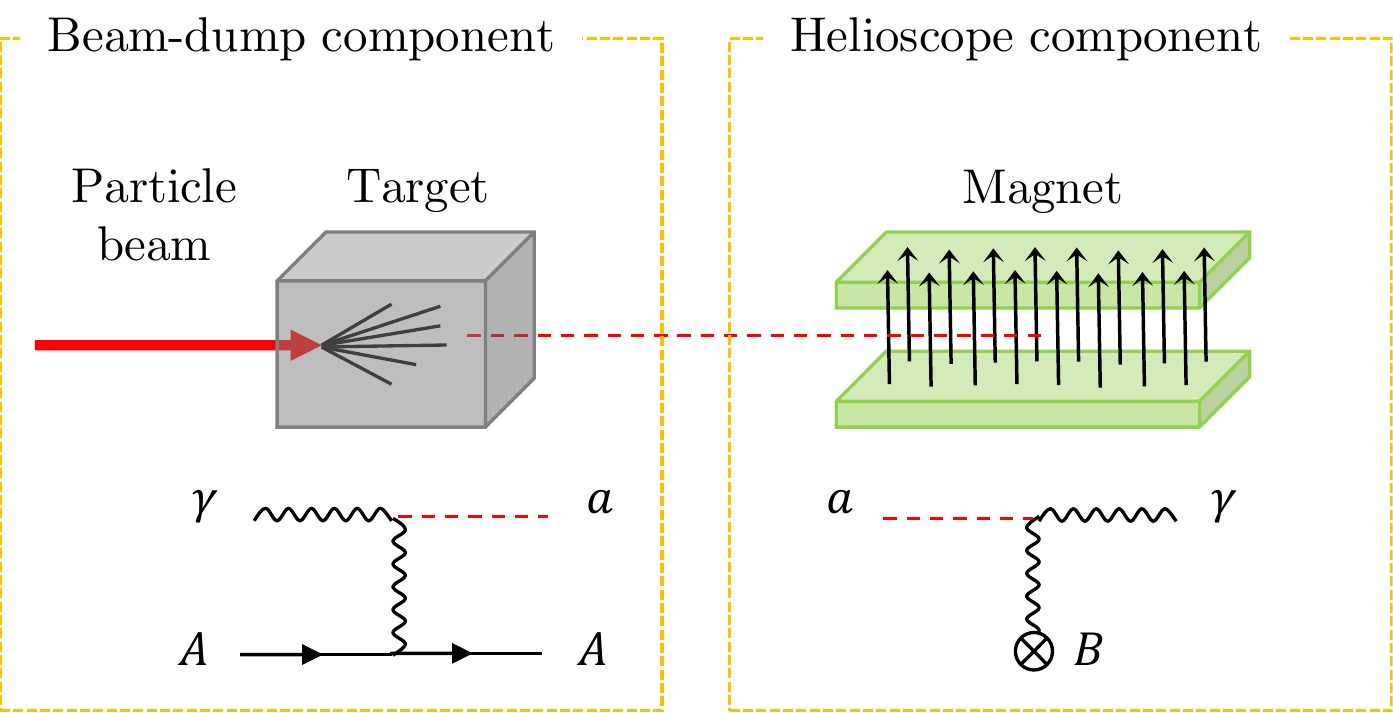}
\caption{\label{fig:prop} A schematic description of the main concept of PASSAT, which is a hybrid of the beam-dump experiment for production of ALP via the Primakoff effect (left) and the helioscope experiment for detection of ALP through the $a\to\gamma$ conversion (right).
 }
\end{figure}

As mentioned in the introduction and also shown in Fig.~\ref{fig:prop}, PASSAT basically consists of two components, production of ALPs as in usual beam-dump experiments and detection of ALPs as in usual helioscope experiments. 
In the beam-dump component, an intense beam of particles (e.g., protons or electrons) impinges on a target, creating a set of particles including photons inside the target. 
A photon can interact with a nearby nucleus and turn into an ALP (henceforth denoted by $a$) via the Primakoff process, i.e., 
\begin{eqnarray}
\gamma+A \to a+A   
\end{eqnarray}
with $A$ symbolizing the atomic system of interest.
In the helioscope component, the produced ALP enters the magnetic field region where it can be converted back to photon which can be readily detected at a conventional photon detector. 

Indeed, wherever a photon is created, an ALP can emerge in the presence of the ALP-photon coupling. Therefore, it is crucial to have a precise estimate of the photon flux in the target in order to estimate the sensitivity reach of a given experiment more precisely~\cite{Dutta:2019nbn,Dent:2019ueq,Dutta:2020vop,Brdar:2020dpr}. 
A few important sources of photons are worth mentioning:  
\begin{itemize}
    \item {\bf Meson decays}: Neutral mesons such as $\pi^0$ and $\eta$ produced in nuclear reactions are good sources of photons, as they promptly decay to SM photons, e.g., $\pi^0 \to \gamma\gamma$, $\eta \to \gamma\gamma$, and $\eta \to 3\pi^0 \to 6\gamma$. 
    \item {\bf Cascade photons}: Primary particles created in the target by beam collision lose their energy by ionizing nearby atoms, hence producing electrons which further go through electromagnetic cascade showers. Charged particles themselves can also radiate off photons on top of the ionization. 
\end{itemize}
In many of the existing studies, only the contributions from meson decays and beam-induced bremsstrahlung photons were considered, resulting in somewhat underestimated experimental sensitivities. 
The sensitivity reaches by the meson contributions can be estimated semi-analytically, using empirical models for meson production in the target. For example, the authors in Ref.~\cite{Bonivento:2019sri} assessed the production of $\pi^0$, adopting the so-called BMPT model~\cite{Bonesini:2001iz} whose original fits were tuned with collision data of protons on a beryllium target.
As is well known, conventional event generators can describe production of mesons in the target. However, additional production of $\pi^0$ and $\eta$ by low-energy effects may not be captured by them. Furthermore, description of cascade photons requires a more dedicated detector-level simulation. 
To this end, we will utilize the \texttt{GEANT4} code package~\cite{Agostinelli:2002hh} to simulate the photon flux in our sensitivity calculations given in Sect.~\ref{sec:result}.

\subsection{Application to DUNE-like Experiments}

In principle, PASSAT can be implemented in any of the beam-dump experiments, either if their detectors have a magnet system or if a decommissioned magnet is installed in an appropriate place. 
For example, the authors in Ref.~\cite{Bonivento:2019sri} investigated the first possibility at the NOMAD experiment as its detector comes with a 0.4~T magnetic field, and studied the second possibility by reusing the magnets of the CAST or the proposed BabyIAXO experiments and placing them at the proposed beam-dump facility at CERN.  

\begin{figure*}[t]
    \centering
    \includegraphics[width=15cm]{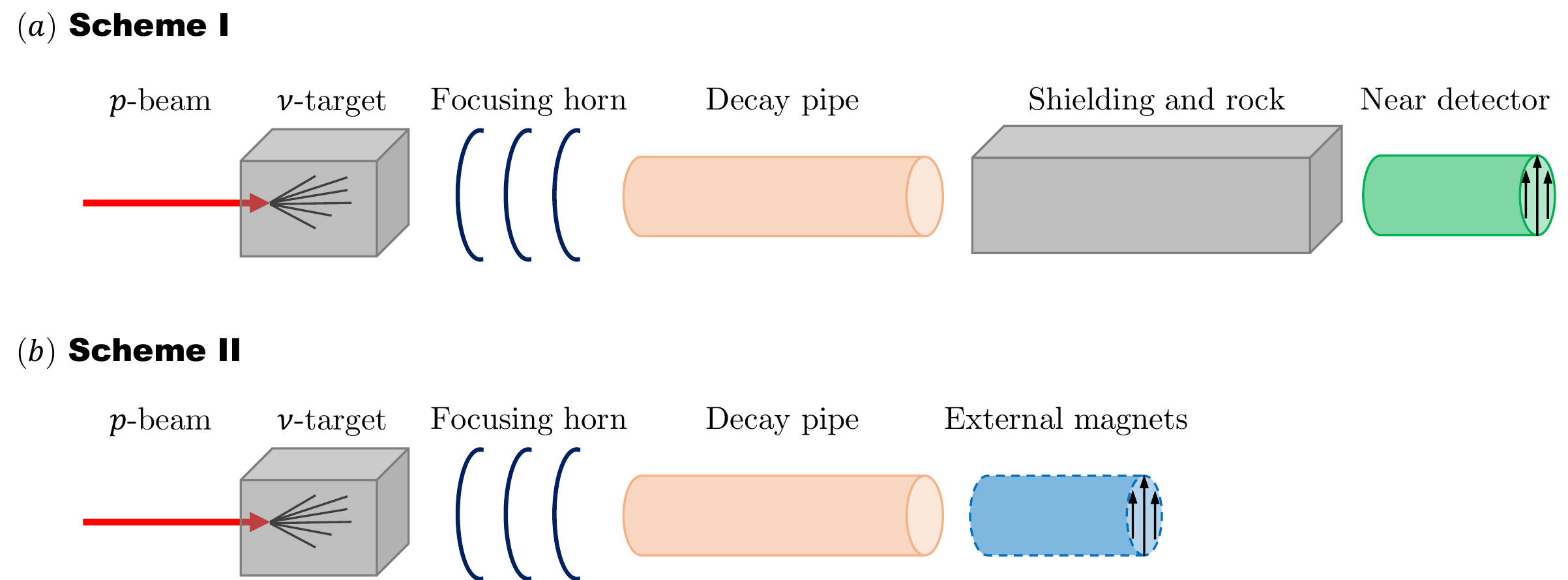}
    \caption{A schematic layout of PASSAT implementation at a typical neutrino oscillation experiment. Scheme I of PASSAT consisting of a proton beam, a neutrino target, and a near detector is possible, if the near detector involves a magnetic field. Scheme II of PASSAT replaces the near detector part by a dedicated magnetic field exerted by external magnets. Here we show the possibility that the magnets are placed downstream from the decay pipe. The external magnets may accompany a photon detector (rear side) and/or shielding material (front side), depending on the detailed experimental design. }
    \label{fig:passatdune}
\end{figure*}

In this paper, we will discuss both possibilities in the beam-produced neutrino oscillation experiments, taking DUNE as a concrete example. 
We sketch the overall configuration of our proposed PASSAT implementation at a typical neutrino beam experiment in Fig.~\ref{fig:passatdune}. 
\begin{itemize}
\item {\bf Scheme I} [Fig.~\ref{fig:passatdune}($a$)]: The first possibility consists of a proton beam, a neutrino target, and a near detector behind the shielding and rock.
Indeed, the near detector complex of DUNE will have the MPD which consists of a high-pressure gas-phase argon time projection chamber (GArTPC) surrounded by an electromagnetic calorimeter in a 0.5~T magnetic field~\cite{Abi:2020wmh}.  
Therefore, our proposed ALP searches by Scheme I can be readily performed with the already existing DUNE setup at no extra cost.
\item {\bf Scheme II} [Fig.~\ref{fig:passatdune}($b$)]: By contrast, the second possibility involves a dedicated magnetic field region. 
As a concrete realization of Scheme II, we conduct a case study of recycling the magnets of the CAST~\cite{Zioutas:1998cc} or the BabyIAXO~\cite{Armengaud:2019uso} experiments and placing them downstream from the decay pipe. The realization of this scheme at DUNE might require some additional civil engineering, but as we show below, this would enable us to gain up to an order of magnitude in ALP sensitivity.
\end{itemize}

The near-detector complex of DUNE will be located 574 meters away from a graphite target. The MPD is part of the near-detector system and will sit immediately downstream of the liquid-phase argon detector.  
The pressure vessel of MPD is of cylindrical shape with diameter and length each being 5~m, of which top and bottom bases are parallel to the beam axis~\cite{Abi:2020evt}.  
In our analysis under Scheme I, we consider a smaller version of MPD-like detector mainly for calculational convenience: a cylinder of 3.2~m diameter and 3.2~m length with the top and bottom bases being beam-axis normal. 
Note that this geometry can be embedded in the fiducial volume of MPD, so our sensitivity estimate for the MPD in Sect.~\ref{sec:result} can be considered somewhat conservative. 
We summarize key experimental parameters of the magnetic field region in Table~\ref{tab:summary}.

\begin{table}[!t]
    \centering
    \caption{Experimental parameters for the magnetic field region of the benchmark detectors. The $B$-field strength of BabyIAXO is the claimed average value. MPD is a MPD-like detector whose geometry can be embedded in the fiducial volume of the DUNE MPD. See text for details. }
    \label{tab:summary}
    \begin{tabular}{c|c c c}
    \hline \hline
    Detector & ~$B$ field [Tesla]~ & ~Length [m]~ & ~Area [m$^2$]~ \\
    \hline
    MPD~\cite{Abi:2020evt} & 0.5 & 3.2 & $8.0$ \\
    CAST~\cite{Zioutas:1998cc} & 8.4 & 9.26 & $1.45\times10^{-3}$ \\
    ~BabyIAXO~\cite{Armengaud:2019uso}~& 2 & 10 & 0.77 \\
    \hline \hline
    \end{tabular}
\end{table}

On the other hand, the decay pipe ends at 221~m from the neutrino target and the (muon) shielding is 49~m away from the decay pipe in the original DUNE near-detector setup. There is a muon alcove between the decay pipe and the shielding~\cite{Abi:2020wmh}, where we propose in our Scheme II to place the magnet from either CAST or BabyIAXO experiment after decommissioning.
The ALPs produced at the neutrino target traverse the focus horns and the decay pipe, and enter the bore of the CAST or BabyIAXO magnet where the ALP-to-photon conversion occurs. 
Key parameters for the magnetic field region of CAST and BabyIAXO magnets are tabulated in Table~\ref{tab:summary}. 
We note that the photon detectors in the CAST and BabyIAXO experiments are sensitive to X-ray-range photons for their original mission. In this study, we assume that the detector is replaced by a suitable $\gamma$-ray photon detector (e.g., an electromagnetic calorimeter). 

\section{Signal Rate \label{sec:rate}}

Since we are interested in the interaction of ALP with photon, we focus on a generic model where the coupling of ALP to photon is described by the following term in the interaction Lagrangian:
\begin{equation}
    \mathcal{L}_{\rm int} \supset -\frac{1}{4}g_{a\gamma\gamma} a F_{\mu\nu}\widetilde{F}^{\mu\nu}\,, \label{eq:lag}
\end{equation}
where $g_{a\gamma\gamma}$ parameterizes the coupling strength and $F_{\mu\nu}$,  $\widetilde{F}_{\mu\nu}$ are the usual field strength tensor of the SM photon and its dual, respectively. 

As mentioned earlier, the signal process begins with a photon created inside the target.
For a given single photon (say, $i$th photon), the probability that an ALP signal event is observed at the detector is essentially given by a product of three relevant probabilities:
\begin{equation}
    P_i = P_{{\rm prod},i} \cdot P_{{\rm surv},i} \cdot P_{{\rm conv},i} \,,
    \label{eq:Pperphoton}  
\end{equation}
where the index $i$ indicates that the probabilities in general differ from one injected photon to another.  
The three probabilities on the right-hand side of Eq.~\eqref{eq:Pperphoton} respectively describe
\begin{itemize}
    \item $P_{\rm prod}$: the probability that a photon undergoes the Primakoff process over the other SM processes and moves to the detector of interest,  
    \item $P_{\rm surv}$: the probability that an ALP traveling toward the detector reaches it without decaying to a photon pair,
    \item $P_{\rm conv}$: the probability that an ALP is converted to a photon by an external magnetic field in the detector.
\end{itemize}
If one takes a sufficiently large set of photons, say $N_\gamma$ photons, one can evaluate the average probability $\langle P \rangle$ 
\begin{equation}
    \langle P \rangle =\frac{1}{N_\gamma} \sum_{i=1}^{N_\gamma} P_i\,.
\end{equation}
This quantity can be understood as the signal acceptance with respect to the injected photons, so the total number of signal events $N_{\rm tot}$ is given by $\langle P \rangle$ multiplied by the total number of photons created in the target $N_{{\rm tot},\gamma}$, i.e.  
\begin{equation}
    N_{\rm tot} = N_{{\rm tot},\gamma} \langle P \rangle\,.
\end{equation}
We now discuss the three individual probabilities in Eq.~\eqref{eq:Pperphoton} one by one below.

\medskip

\noindent {i) $P_{\rm prod}$}: In the presence of the interaction in Eq.~\eqref{eq:lag}, a photon produced in the target is either converted to an ALP via the Primakoff process or absorbed through standard interactions, such as pair production, photoelectric absorption, etc. Therefore, we have
\begin{equation}
\label{eqn:Pprod}
    P_{\rm prod}=\frac{1}{\sigma_{\rm SM}+\sigma_P} \int d\theta'_a d\phi'_a \frac{d^2\sigma_P}{d\theta'_ad\phi'_a}\,,
\end{equation}
where $\theta'_a$ and $\phi'_a$ are respectively the polar and azimuthal angles of the outgoing ALP with respect to the direction of the incoming photon.
The differential Primakoff production cross-section is integrated over all angular ranges within which the produced ALP can pass through the detector (see Appendix~\ref{app:coordinates} for more calculational details). 
$\sigma_P$ is the total production cross-section, while $\sigma_{\rm SM}$ denotes the total cross-section of standard interactions mentioned above Eq.~(\ref{eqn:Pprod}), which is, in general, a function of the photon energy~\cite{xcom}. 
If the photon energy is large enough, $\sigma_{\rm SM}$ does not vary much. However, in our analysis, we will plug $\sigma_{\rm SM}=\sigma_{\rm SM}(E_\gamma)$ into $P_{\rm prod}$ for more precise estimates, based on the measurement data in Ref.~\cite{xcom}. 
The differential production cross-section with respect to the angles $\theta'_a$ and $\phi'_a$ is well known~\cite{Tsai:1986tx}, which is given by  
\begin{equation}
    \frac{d^2\sigma_P}{d\theta'_a d\phi'_a}=\frac{1}{8\pi}g_{a\gamma\gamma}^2\alpha Z^2 \left[F(t)\right]^2 \frac{p_a^4 \sin^3 \theta'_a}{t^2}\,, \label{eq:prodxs}
\end{equation}
where $\alpha$, $p_a$, and $Z$ are the electromagnetic fine structure constant, ALP momentum, and the atomic number of the target material, respectively, and $t$ denotes the squared momentum transfer which is given by 
\begin{eqnarray}
t  =  m_a^2-2E_\gamma(E_a-p_a\cos\theta'_a)\,, 
\end{eqnarray}
$E_a$ being the ALP energy. 
Finally, $F(t)$ in Eq.~\eqref{eq:prodxs} describes the form factor. Considering the typical energy scale of photons in our analysis in Sect.~\ref{sec:result}, we use the nuclear form factor with the Helm parameterization:
\begin{equation}
    F(t)=\frac{3j_1(\sqrt{|t|}R_1)}{\sqrt{|t|}R_1}\exp\left(-\frac{|t|s^2}{2} \right)\,,
\end{equation}
where $j_1$ is the spherical Bessel function of the first kind. Here $s=0.9$~fm and $R_1$ is parameterized as per Ref.~\cite{Lewin:1995rx}, i.e. $R_1 = \sqrt{(1.23A^{1/3}-0.6)^2+2.18}~\text{fm}$, with $A$ being the atomic mass number of the target material. 

\begin{figure}[t]
    \centering
    \includegraphics[width=8.4cm]{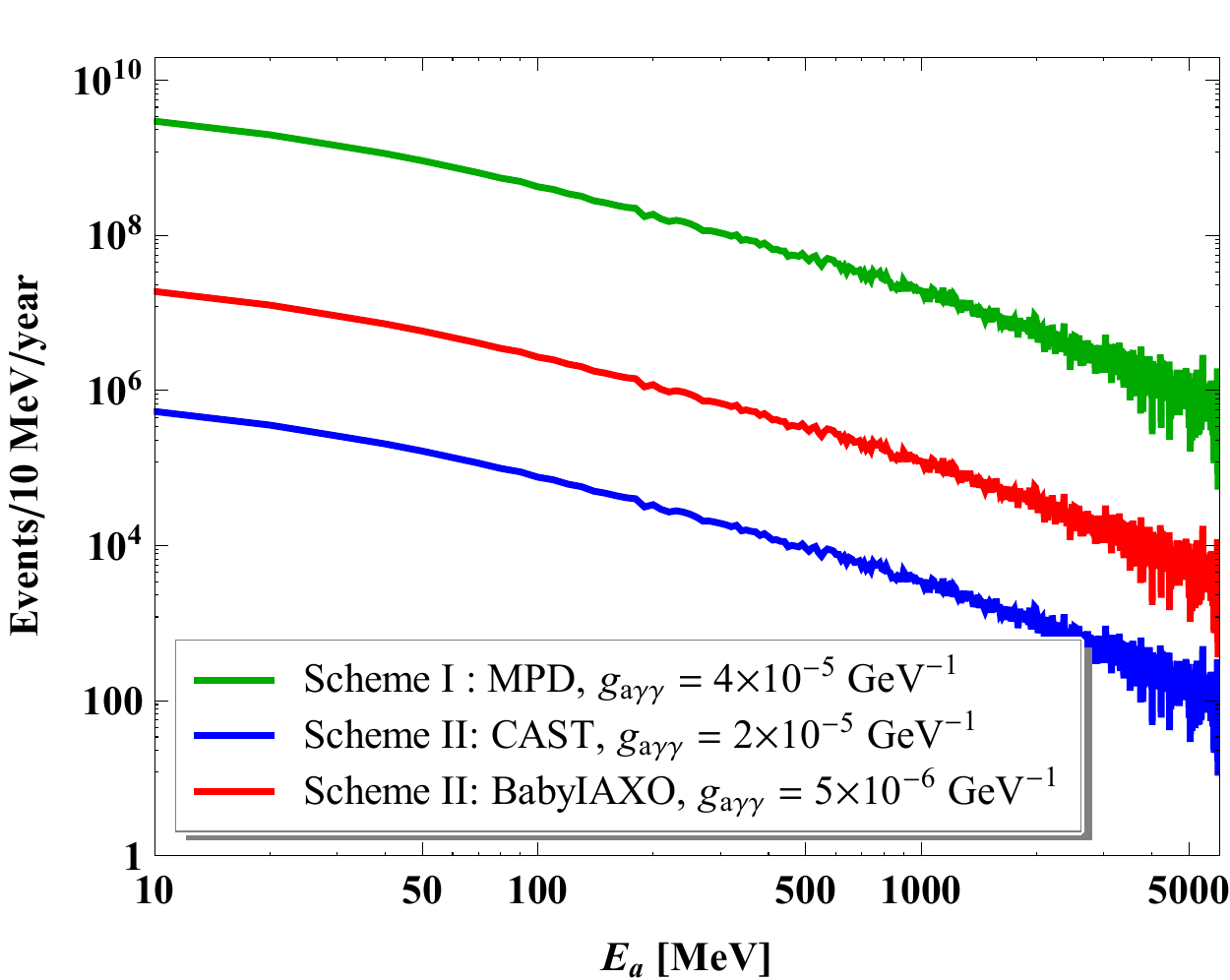}
    \caption{Energy spectra of 1-eV ALPs produced per year at the target. Only selected are ALPs which would travel toward the three benchmark detector systems, Scheme I with the MPD (green) and Scheme II with the magnets of CAST (blue) or BabyIAXO (red). 
    The couplings in the legend are the values at which sensitivity arises (see Fig.~\ref{fig:expsense}). The fluctuations at higher $E_a$ are due to low  statistics. }
    \label{fig:ALPenergy}
\end{figure}

We will work on the Primakoff production cross-section in the collinear limit, as the momentum transfer to the target nucleus is much smaller than the energy $E_\gamma$ of photons created in the target. As a result, the energy of ALP $E_a$ is almost the same as the photon energy as far as it is greater than the ALP mass, i.e., $E_a\approx E_\gamma (> m_a)$.
We display energy spectra of ALPs produced per year at the target in Fig.~\ref{fig:ALPenergy}. Here only selected are ALPs which would fly to the detectors of the three benchmark scenarios, Scheme I with the MPD (green) and Scheme II with the magnets of CAST (blue) or BabyIAXO (red). We choose $m_a=1$~eV as an illustration; the spectral shapes do not depend on the choice of $m_a$  as long as $E_\gamma \gg m_a$. 
The couplings in the legend are set to be the values at which sensitivity arises, in order to develop the intuition of required ALP fluxes (see Fig.~\ref{fig:expsense}).  
The required ALP flux for the MPD case is larger than the CAST and BabyIAXO cases as its effective detector length and its strength of the magnetic field are smaller than those of the others.

\medskip

\noindent {ii) $P_{\rm surv}$}: Following the usual decay law, the survival probability has the form of
\begin{equation}
    P_{\rm surv}=\exp\left(-\frac{\ell}{\ell_a^{\rm lab}}\right)\,,
\end{equation}
for a given distance $\ell$ between the target and the detector of interest. Here $\ell_a^{\rm lab}$ denotes the laboratory-frame mean decay length of ALP which is a function of the decay width of $a\to 2\gamma$, i.e. 
\begin{equation}
    \Gamma_a =\frac{g_{a\gamma\gamma}^2m_a^3}{64\pi}\,, 
\end{equation}
and the Lorentz boost factor of the decaying ALP $\gamma_a$:
\begin{eqnarray}
    \ell_a^{\rm lab}&=&\frac{\sqrt{\gamma_a^2-1}}{\Gamma_a} \nonumber \\
    &\approx& 4,000~{\rm m}\left(\frac{\gamma_a}{10^2} \right)^2 \left(\frac{10^{-3}~{\rm GeV}^{-1}}{g_{a\gamma\gamma}} \right)^2\left( \frac{1~{\rm MeV}}{m_a} \right)^3 . \nonumber \\ &&
\end{eqnarray}

\medskip 

\noindent {iii) $P_{\rm conv}$}: When an ALP travels a distance $L$ in a magnetic field $B$, the ALP$-$photon conversion probability is~\cite{Irastorza:2018dyq}
\begin{equation}
    P_{\rm conv}=\left(\frac{g_{a\gamma\gamma}BL}{2}\right)^2 \left(\frac{qL}{2} \right)^{-2} \sin^2 \left( \frac{qL}{2}\right)\,, \label{eq:pconv}
\end{equation}
where the last two factors reflect the coherence of the conversion. 
In the relativistic limit and in vacuum, the quantity $q$ is expressed as
\begin{equation}
    q=2 \sqrt{\left(\frac{m_a^2}{4E_a} \right)^2 +\left(\frac{1}{2}g_{a\gamma\gamma}B \right)^2}\,.
\end{equation}
In the limit of $g_{a\gamma\gamma} \to 0$ and $m_a\to0$, we have $qL\ll 1$, and then the conversion probability effectively becomes $m_a$-independent and can be approximated to be 
\begin{eqnarray}
    P_{\rm conv}&\approx& \left(\frac{g_{a\gamma\gamma}BL}{2}\right)^2 \nonumber \\ 
    &=& 2.45\times10^{-11} \left( \frac{g_{a\gamma\gamma}}{10^{-5}~{\rm GeV}^{-1}} \frac{B}{1~{\rm T}} \frac{L}{1~{\rm m}}\right)^2. 
\end{eqnarray}

Note that this ALP-to-photon conversion process does not change the photon energy. Therefore, the photon from the conversion, $\gamma_{\rm conv}$ has the same magnitude of energy as the incoming ALP, i.e., $E_{\gamma_{\rm conv}}=E_a$.

\section{Background Consideration \label{sec:background}}

In this section, we discuss potential backgrounds for our ALP signal. 
Since a signal event involves a photon in the final state, none of the known SM processes would result in a single photon. 
However, once realistic detector effects such as threshold and detection efficiencies are taken into account, several SM processes would behave like signal events, giving rise to a large number of background events. 
Nevertheless, in our analysis in Sect.~\ref{sec:result}, we assume $\mathcal{O}(100)$ and negligible background events in Scheme I and Scheme II, respectively. 
We now argue that this level of background reduction is indeed achievable, while we reserve a dedicated background simulation study for future work. 

For Scheme I, potential backgrounds are from the neutrino neutral-current (NC) single $\pi^0$ events, for example,
\begin{eqnarray}
\nu+n &\to& \nu+n+\pi^0 \,, \\
\nu +A &\to& \nu + A+\pi^0 \,,
\end{eqnarray}
in which one of the two photons from the $\pi^0$ decay is undetected in the GArTPC of MPD. 
The photon conversion probability in the gas is about 12\%~\cite{Dutta:2020vop}, so 21\% of $\pi^0$'s would appear single photon-like. 
We estimate the rate of these events, combining the $\nu_\mu$/$\bar{\nu}_\mu$ and $\nu_e$/$\bar{\nu}_e$ fluxes as reported in Ref.~\cite{Abi:2020evt} with the scattering cross-sections of $\nu+n \to \nu+n+\pi^0$ (resonance $\pi^0$ production) and $\nu +A \to \nu + A+\pi^0$ (coherent $\pi^0$ production) from Ref.~\cite{Formaggio:2013kya}. We find that $\sim 10^5$ such events would be identified as potential background events, assuming a 1-ton fiducial mass and a 7-year exposure (3.5 years in the neutrino mode $+$ 3.5 years in the antineutrino mode). 

Kinematic cuts can significantly suppress these background events. 
The ALP signal flux basically comes from the target, and therefore the momentum direction of the converted photons lies within about 2.7~mrad from the beam axis inside the MPD. 
By contrast, the $\pi^0$'s in the background events are less forward-scattered. 
The Monte Carlo study in Ref.~\cite{Brdar:2020dpr} estimated that imposing a 4~mrad angular cut would suppress these backgrounds by a factor of $10^3$, leaving 100 signal-faking events for a 7-year exposure.  
We remark that a harder angular cut can reduce the backgrounds further by a factor of few, as there is a possibility that the final angular resolution of MPD would be as small as 2~mrad~\cite{Abi:2020wmh}.
However, we conservatively assume 100 background events in our sensitivity calculations for MPD.

On the other hand, for Scheme II, background events are produced in different ways. We closely follow the estimates in Ref.~\cite{Bonivento:2019sri}. As mentioned in Sect.~\ref{sec:idea}, a photon detector designed to be sensitive to $\gamma$-ray photons (e.g., calorimeter) is needed. Therefore, an irreducible background will arise from $\nu$-$e^-$ elastic scattering events inside the photon detector. Assuming the same photon detector specification as described in Ref.~\cite{Bonivento:2019sri}, we expect $\mathcal{O}(20)$ background events.
However, this background emerges irrespective of the presence of the magnetic field. Therefore, an {\it in-situ} estimation is possible by running an experiment with the magnet turned off, and the background can be essentially removed.\footnote{This  is  true  only  up  to  some  statistical error. Precise estimates and the associated uncertainties in this way of background subtraction depend on the fraction of runtime to be devoted for the off-magnet mode, as well as on the error propagation. These details are beyond the scope of this study, and we leave the detailed experimental strategy to interested experimental collaborations.} 

In addition, a large flux of $\mu^\pm$ and $e^\pm$ will still reach the CAST or BabyIAXO magnets due to the high proton beam intensity. To suppress backgrounds induced by these charged particles, a set of appropriately optimized veto detectors and shielding material will be needed upstream from the magnets. Also, to avoid potential interactions (e.g., scattering) of particles inside the bore region of the CAST or BabyIAXO module, a certain level of vacuum could be necessary while a detailed design proposal is beyond the scope of this paper. 
Again, their occurrence has nothing to do with the magnetic field, so they can in principle be removed by the above-described {\it in-situ} background estimation. 

Given these considerations, we take a negligible background hypothesis in our sensitivity calculations under Scheme II, while even assuming $\mathcal{O}(20)$ ``irreducible'' backgrounds would not yield an appreciable change in the sensitivity reaches due to $\sim g_{a\gamma\gamma}^4$ dependence (two powers from production and the other two powers from detection).

\section{Simulations and Results \label{sec:result}}

Our analysis begins with a photon flux generation inside the target of a DUNE-like experiment. 
We take a 1.5-meter long graphite target of a simple cylindrical shape, and make use of a \texttt{GEANT4} simulation sample~\cite{Brdar:2020dpr}.
The production cross-section of ALPs via the Primakoff effect in Eq.~\eqref{eq:prodxs} is calculated in the collinear limit, i.e., $E_\gamma\approx E_a$, as the momentum transfer to the target nucleus is much smaller than the energy $E_\gamma$ of photons created in the target. In this limit, the energy of the converted photons in the detector is equal to $E_\gamma$ since $E_a$ is essentially transferred to $E_{\gamma_{\rm conv}}$ as mentioned before. We then have $E_\gamma \approx E_{\gamma_{\rm conv}}$, and we consider photons of energy $E_\gamma>5$~MeV in Scheme I, following the claimed energy threshold of the DUNE MPD~\cite{Abi:2020wmh}.
In addition, a 12\% photon pair-conversion probability in the gas-phase detector~\cite{Dutta:2020vop}, as mentioned earlier, is taken as the baseline photon detection efficiency.
In Scheme II, the corresponding cut depends on the photon detector specifications. Here we take a 5 MeV threshold for simplicity. 

\begin{figure*}[t!]
\centering
\includegraphics[width=0.8\textwidth]{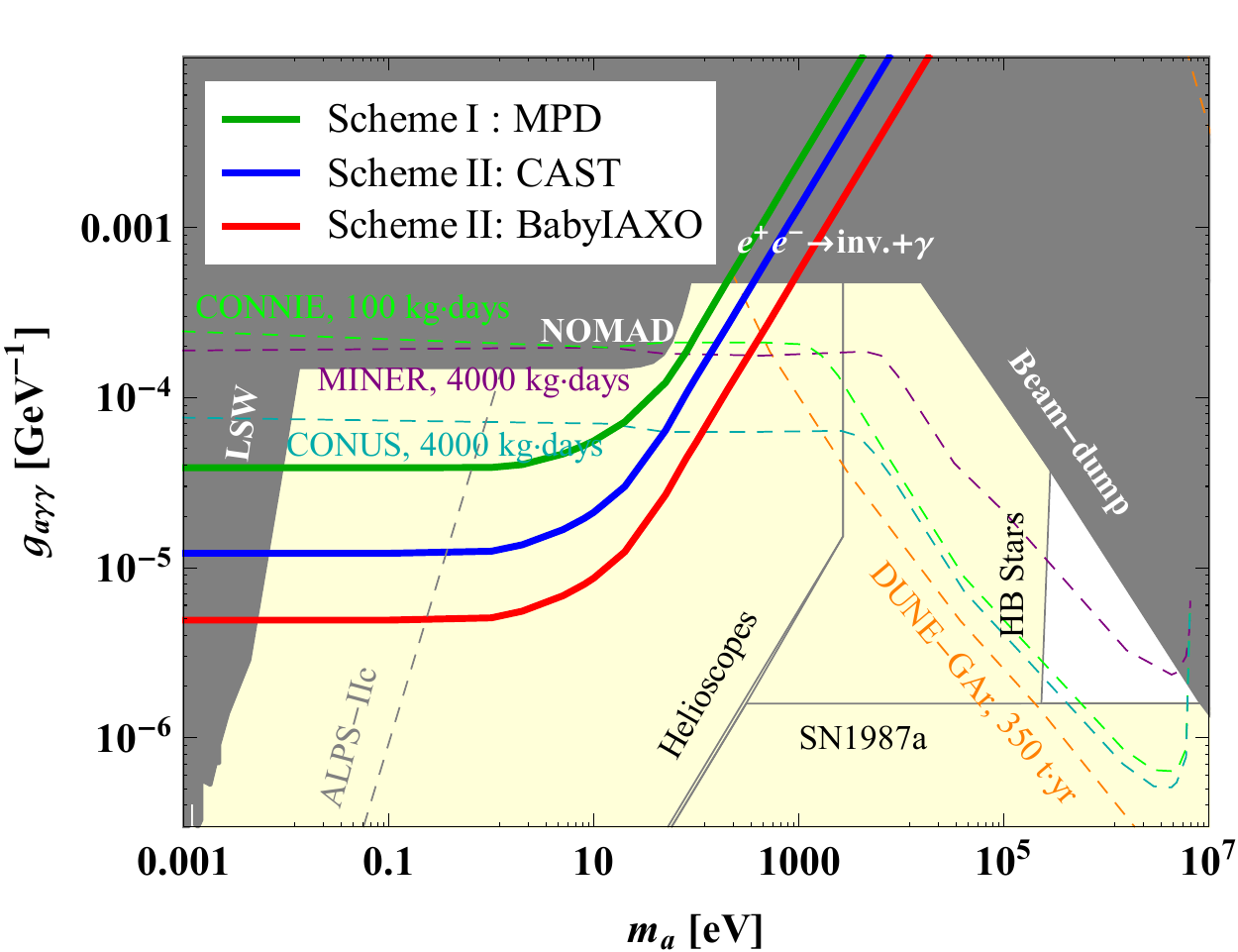}
\caption{\label{fig:expsense} Expected sensitivity reaches that can be achieved by an implementation of the PASSAT idea in a DUNE-like experiment according to Scheme I with the MPD (green) and Scheme II with the magnets of CAST (blue) or BabyIAXO (red), in the plane of ALP mass $m_a$ and the the associated photon coupling $g_{a\gamma\gamma}$. The limits are estimated at 90\% C.L., under the assumptions of 100 (negligible) background events for Scheme I (Scheme II) with an exposure of $1.1\times 10^{21}$ POT/year $\times$ 7 years. The shaded regions represent the existing constraints summarized in Refs.~\cite{Bauer:2018uxu,Lanfranchi:2020crw}: the gray regions are excluded by existing laboratory-produced ALP searches, while those in light yellow show the (potentially avoidable) exclusion limits from existing astrophysical ALP searches. 
For comparison purposes, we also show the future sensitivity reaches expected at the gaseous argon near-detector of DUNE in the decay channel (orange dashed line)~\cite{Brdar:2020dpr} and at reactor neutrino experiments MINER (purple dashed line), CONNIE (green dashed line), and CONUS (cyan dashed line), in the scattering (horizontal dashed lines) and decay channels~\cite{Dent:2019ueq}. The final-stage ALPS-II expectation (ALPS-IIc)~\cite{Bahre:2013ywa} is also shown by the dashed gray line.
}
\end{figure*}

The expected sensitivity reaches at 90\% C.L. in the $(m_a, g_{a\gamma\gamma})$ plane are shown in Fig.~\ref{fig:expsense}. 
We evaluate the limits, taking into account statistical fluctuation associated with 100 and negligible background events in Scheme I (green for MPD) and Scheme II (blue for CAST and red for BabyIAXO), respectively, for a 7-year data collection (3.5 years in the neutrino mode $+$ 3.5 years in the antineutrino mode). 
Various existing constraints compiled in Refs.~\cite{Bauer:2018uxu,Lanfranchi:2020crw} are also shown in Fig.~\ref{fig:expsense}. Specifically, the gray regions are excluded by the laboratory-produced ALP searches, with the boundaries set by LSW (e.g., ALPS-I), NOMAD, electron-positron colliders, and beam-dump experiments (e.g., E137). For reference purposes, the regions constrained by astrophysical ALP searches are shown in light yellow (see also Ref.~\cite{Lucente:2020whw} for a recent development in the supernova limit calculation and Ref.~\cite{Bar:2019ifz} for its criticism). As mentioned in Sect.~\ref{sec:intro}, these astrophysical limits can  potentially be evaded, so we consider them as less robust than the laboratory ones. 
Also, standard cosmological considerations would constrain the white region, the so-called ``cosmological triangle''. Like the astrophysical limits, however, they depend highly on underlying model details, hence can be evaded by non-standard cosmology~\cite{Carenza:2020zil,Depta:2020wmr}.

The green line in Fig.~\ref{fig:expsense} suggests that the DUNE MPD itself is capable of probing $g_{a\gamma\gamma}$ as small as a few times $10^{-5}$ for $m_a\lesssim 10$~eV, setting a new laboratory-based limit beyond the existing ones. 
In this mass range, production and conversion of ALPs are essentially independent of the mass of ALP because $E_\gamma \gg m_a$ in the target and $qL \ll 1$ in the MPD, respectively. 
Beyond $m_a \sim10$~eV, the decoherence encoded in the last two factors of Eq.~\eqref{eq:pconv} becomes gradually substantial with increasing $m_a$, i.e., the conversion probability essentially vanishes in the limit of large ALP mass. 

To discuss the complementarity between this conversion channel and other search channels, we show in Fig.~\ref{fig:expsense} the relevant future sensitivity reaches expected at the gaseous argon near-detector of DUNE in the decay channel (orange dashed line)~\cite{Brdar:2020dpr} and at reactor neutrino experiments MINER (purple dashed line), CONNIE (green dashed line), and CONUS (cyan dashed line), in the scattering (horizontal dashed lines) and decay channels~\cite{Dent:2019ueq}. It is clear that in the small ALP mass regime, the PASSAT implementation utilizing the ALP-photon conversion channel is expected to give sensitivity comparable to or better than that of the scattering channel, whereas in the large ALP mass limit, the decay of ALP to two photons is a more effective probe than the ALP-photon conversion.

The work in Ref.~\cite{Brdar:2020dpr} showed that the liquid argon near-detector of DUNE can also be sensitive to the ALP scattering signal because scattering limits get better with more scattering targets, although it would suffer from more backgrounds. 
Therefore, the approach presented in our Scheme I provides complementary information in probing the ALP parameter space, allowing a DUNE-like experiment to be equipped with a complete set of ALP search channels, i.e. ALP-to-photon conversion in addition to the ALP decay and scattering channels investigated in Ref.~\cite{Brdar:2020dpr}. 

In addition, we show in Fig.~\ref{fig:expsense} the final-stage ALPS-II expectation (ALPS-IIc)~\cite{Bahre:2013ywa} by the dashed gray line, in order to compare our results with future limits of other laboratory-produced ALP searches, LSW-type experiments in particular. 
As shown here, the ALPS-IIc prediction would be competitive toward smaller couplings with sub-eV mass or below. 
Our proposal shows competitive sensitivity in different regions of parameter space, providing complementarity in the search for laboratory-produced ALPs. 

When it comes to Scheme II, where we propose reusing the magnets of CAST or BabyIAXO, we expect that the sensitivity reaches will be improved by at least a factor of few and up to an order of magnitude, compared to that of MPD at a DUNE-like experiment. 
The reason is that the magnets are placed closer to the target and their magnetic field strengths are larger than that of MPD. 
It is interesting that the case with CAST can have competitive sensitivity reaches despite its small bore. The photons emerging in the target are forward-directed, and, in turn, the ALP flux is forward-directed so that a large fraction of ALPs can enter the magnetic field region even through the small aperture. 

Finally, we comment on the position of the external magnets. As described earlier, it would be ideal to have additional shielding material in front of the magnets mainly to suppress a large flux of beam remnants and beam-induced (charged) particles. 
Depending on the neutrino experiment of interest, there may be no adequate space behind the decay pipe, and thus one may place the magnets behind the existing shielding or dump area instead of attaching the shielding material to the magnets. 
This may require a certain level of civil engineering to secure the space for the magnets. 
Establishing cost-efficient and/or feasible experimental designs is beyond the scope of this paper, so we leave this task to the experimental collaborations. The physics gain of this proposed Scheme II is clear from Fig.~\ref{fig:expsense} for the ALP case. But depending on the community interest, it might be worth exploring other physics potentials of this design scheme with a powerful magnet at the near-detector site.

\section{Conclusion \label{sec:conclusion}}

We have discussed possible implementations of the idea of PASSAT proposed in Ref.~\cite{Bonivento:2019sri} in future beam neutrino experiments, and have investigated the parameter space of ALP at a DUNE-like experiment, using the ALP-photon coupling.  
Since the MPD in the near detector complex comes with a magnetic field, DUNE can realize PASSAT on its own without any additional cost. We have further proposed an alternative realization by recycling the magnets of CAST or BabyIAXO (after they are decommissioned) and placing them behind the decay pipe.

ALPs are produced in the target via the Primakoff process of photons. For a more precise assessment of sensitivity reaches, we have utilized a \texttt{GEANT4} simulation output to estimate a photon flux, from which we estimate the fraction of ALP flux entering the magnetic field region where ALPs are converted back to photons. With potential backgrounds taken into consideration, our simulations suggest that a DUNE-like experiment implementing PASSAT can probe a wide range of parameter space that none of the laboratory-produced ALP search experiments have explored so far. As shown in Fig.~\ref{fig:expsense}, for ALP mass $m_a \lesssim 10$ eV, the sensitivity at MPD can reach up to few times $10^{-5}$, which can be further improved by almost one order of magnitude by PASSAT realized with the CAST or BabyIAXO magnets. In particular, the expected sensitivities in Fig.~\ref{fig:expsense} cover regions investigated by the CAST helioscope experiment, providing a complementary and robust laboratory probe. 

\section*{Acknowledgements}
We would like to thank Joshua Barrow, Bhaskar Dutta, Ian Shoemaker, and Jae Yu for useful discussions. 
DK particularly appreciates Wooyoung Jang for his \texttt{GEANT4} simulations. The work of BD is supported in part by the US Department of Energy under Grant No. DE-SC0017987, by the Neutrino Theory Network Program Grant No. DE-AC02-07CHI11359, and by a Fermilab Intensity Frontier Fellowship. The work of DK is supported by DOE under Grant No. DE-FG02-13ER41976/DE-SC0009913/DE-SC0010813. The work of KS is supported by U.S. Department of Energy grant number~DE-SC0009956.

\appendix

\section{Coordinates for ALP production 
\label{app:coordinates}}

\begin{figure}[t]
    \centering
    \includegraphics[width=8cm]{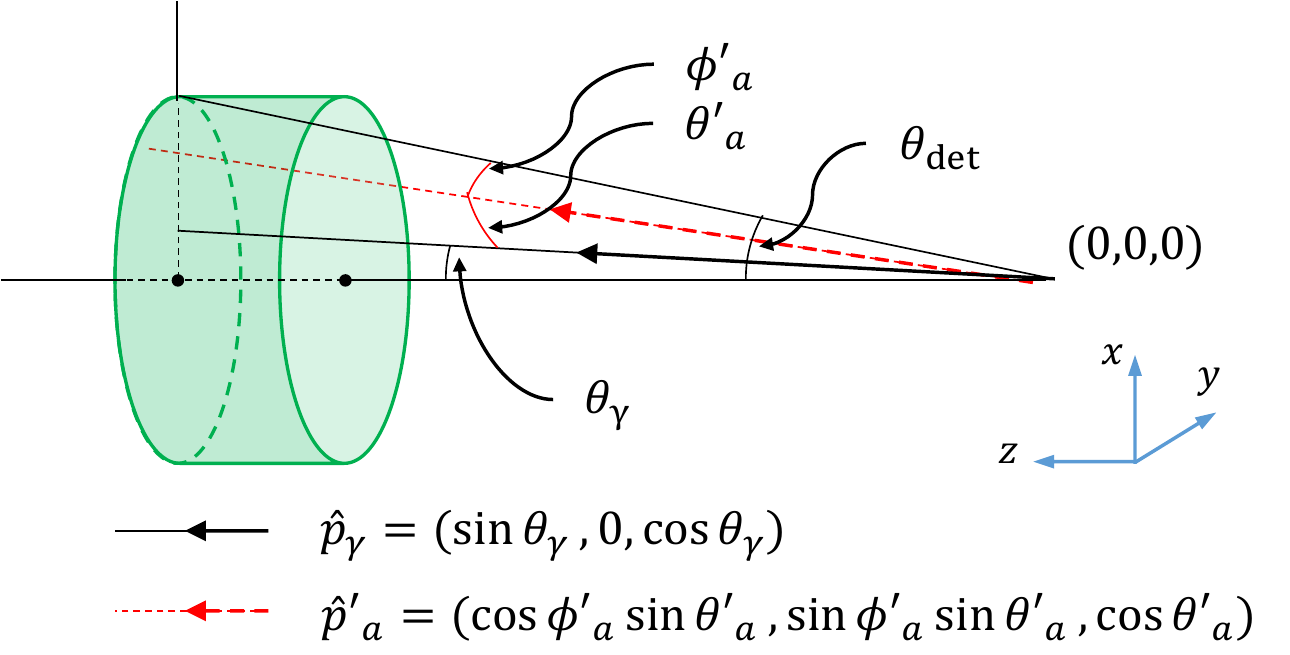}
    \caption{A configuration of an incoming photon, an outgoing ALP, and a detector (green shaded area). The photon is assumed to emerge at $(0,0,0)$. The unprimed angles are defined with respect to the horizontal beam axis, while the primed angles are defined with respect to the incoming photon direction.}
    \label{fig:detconfig}
\end{figure}

For a more systematic discussion on the integration limits, we define a few relevant coordinates and take a cylindrically symmetric detector, as displayed in Fig.~\ref{fig:detconfig}. Let us suppose that a photon emerges at $(0,0,0)$ with an angle $\theta_\gamma$ from the beam axis.
Then the photon direction $\hat{p}_\gamma$ is given by 
\begin{eqnarray}
\hat{p}_\gamma \ = \ (\sin\theta_\gamma,0,\cos\theta_\gamma) \,.
\end{eqnarray}
The outgoing ALP direction with respect to $\hat{p}_\gamma$ is 
\begin{eqnarray}
\hat{p}'_a \ = \ (\cos\phi'_a\sin\theta'_a,\sin\phi'_a\sin\theta'_a,\cos\theta'_a)  \,.
\end{eqnarray}
One can find that the polar angle of ALP with respect to the beam axis $\theta_a$ is
\begin{equation}
    \theta_a = \cos^{-1} (\cos\theta_a^\prime \cos\theta_\gamma + \cos\phi_a^\prime \sin\theta_a^\prime \sin\theta_\gamma)\,.
    \label{eq:angle}
\end{equation}
If the detector covers up to $\theta_{\rm det}$ as depicted in Fig.~\ref{fig:detconfig}, any ALP with $\theta_a \leq \theta_{\rm det}$ can traverse the magnetic field region. The maximum and minimum $\theta'_a$ satisfying this requirement are
\begin{equation}
\left\{
\begin{array}{l l}
    \theta'_{a,\max} &= \theta_\gamma +\theta_{\rm det} \\ [0.5em]
    \theta'_{a,\min} &= \max(0,\theta_\gamma-\theta_{\rm det}),
    \end{array}\right.
\end{equation}
as far as $\theta'_{a,\max}<\pi$. Once $\theta'_{a,\max}$ becomes larger than $\pi$, the above relations should be replaced by
\begin{equation}
    \left\{
    \begin{array}{l l}
    \theta'_{a,\max} &= \pi \\ [0.5em]
    \theta'_{a,\min} &= \min(\theta_\gamma-\theta_{\rm det}, 2\pi-\theta_\gamma-\theta_{\rm det})
    \end{array}\right. \hbox{ for }\theta_\gamma+\theta_{\rm det}>\pi.
\end{equation}
Next, one can find that for a given $\theta'_a$, the $\theta_a \leq \theta_{\rm det}$ condition yields
\begin{eqnarray}
    | \phi'_a| &\leq& \cos^{-1}\left( \max\left[-1,\frac{\cos\theta_{\rm det}-\cos\theta'_a\cos\theta_\gamma}{\sin\theta'_a\sin\theta_\gamma}\right] \right) \nonumber \\
    &\approx& \cos^{-1}\left( \max \left[-1, \frac{\theta_a^{\prime 2} + \theta_\gamma^2 - \theta_{\rm det}^2 }{2\theta'_a \theta_\gamma} \right] \right),
\end{eqnarray}
where the second line is valid for $\theta'_a, \theta_\gamma,\theta_{\rm det} \ll 1$.

\bibliography{main}
\bibliographystyle{JHEP}

\end{document}